%% file: electricity_price_forecasting.tex
\documentclass[authoryear,review,11pt]{elsarticle}

\usepackage{amssymb}

\usepackage{endnotes}
\let\footnote=\endnote

\usepackage[top=2.5cm,left=2.5cm,right=2.5cm,bottom=2.5cm]{geometry}

\usepackage{eurosym}
\usepackage{siunitx}
    \DeclareSIUnit\eur{\officialeuro}
    \DeclareSIUnit\M{M}
    \DeclareSIUnit\k{k}
    \newcommand{\sym}[1]{\rlap{$^{#1}$}}
\usepackage{csquotes}
\usepackage{amsmath}
\usepackage{xspace}
	\newcommand\ie{i.\,e.\xspace}
	\newcommand\eg{e.\,g.\xspace}

	\newcommand\etc{etc.\xspace}

	\newcommand\US{U.\,S.\xspace}

\usepackage{url}

\usepackage{isomath}
\usepackage{bm}
\usepackage[amsmath,thmmarks,hyperref]{ntheorem}
  \theoremstyle{plain}

\DeclareMathOperator*{\argmax}{arg\,max}
\DeclareMathOperator*{\argmin}{arg\,min}

\newcommand{\mathup}[1]{\mathrm{#1}}

	\newcommand{\sspace}{\quad}
		
	\newcommand{\sand}{\sspace\text{and}\sspace}

  \newcommand\define{\ensuremath{\mathrel{\stackrel{\mathrm{def}}{=}}}}
  \newcommand{\abs}[1]{\left\lvert #1 \right\rvert}

  \newcommand{\norm}[1]{\left\lVert#1\right\rVert}

\usepackage[usenames,dvipsnames,table]{xcolor}

\usepackage{algorithm}
\usepackage{algpseudocode}
 
\usepackage{tabularx}
\usepackage{booktabs}
\usepackage{dcolumn}
\usepackage{ragged2e}
  \newcommand{\tablehead}{\bfseries}
\newcommand{\PreserveBackslash}[1]{\let\temp=\\#1\let\\=\temp}
\newcolumntype{v}[1]{>{\PreserveBackslash\RaggedRight\hspace{0pt}}p{#1}}

  \usepackage{adjustbox}
\newcolumntype{R}[2]{%
    >{\adjustbox{angle=#1,lap=\width-(#2)}\bgroup}%
    l%
    <{\egroup}%
}
\newcommand*\rottblhead{\multicolumn{1}{R{45}{1em}}}

\newcommand{\lcellt}[2][l]{%
  \begin{tabular}[t]{@{}#1@{}}#2\end{tabular}}

\usepackage[
    colorinlistoftodos,
    textsize=footnotesize,
        ]{todonotes}
        
\makeatletter
    \renewcommand{\fps@figure}{H}         
    \renewcommand{\fps@table}{H}         
\makeatother 

\usepackage{float}

 \biboptions{comma}

\journal{arXiv}

\begin{document}

\begin{frontmatter}



\title{Forecasting electricity prices with machine learning: Predictor sensitivity}

\author[ETH]{Christof Naumzik\corref{cor1}}
\ead{cnaumzik@ethz.ch}
\author[ETH,Freiburg]{Stefan Feuerriegel}
\ead{sfeuerriegel@ethz.ch}
\address[ETH]{ETH Zurich, Weinbergstr. 56/58, 8092 Zurich, Switzerland}
\address[Freiburg]{Chair for Information Systems Research, University of Freiburg, Platz der Alten Synagoge, 79098 Freiburg, Germany}
\cortext[cor1]{Corresponding author.}

\begin{abstract}

\noindent
\emph{Purpose}: Trading on electricity markets occurs such that the price settlement takes place before delivery, often day-ahead. In practice, these prices are highly volatile as they largely depend upon a range of variables such as electricity demand and the feed-in from renewable energy sources. Hence, accurate forecasts are demanded. 

\noindent
\emph{Approach}: This paper aims at comparing different predictors stemming from supply-side (solar and wind power generation), demand-side, fuel-related and economic influences. For this reason, we implement a broad range of non-linear models from machine learning and draw upon the information-fusion-based sensitivity analysis.

\noindent
\emph{Findings}: We disentangle the respective relevance of each predictor. We show that external predictors altogether decrease root mean squared errors by up to 21.96\%. A Diebold-Mariano test statistically proves that the forecasting accuracy of the proposed machine learning models is superior.

\noindent
\emph{Originality}: The benefit of adding further predictors has only recently received traction; however, little is known about how the individual variables contribute to improving forecasts in machine learning. 

\noindent
\emph{Research implications}: The performance gain from including more predictors might be larger than from a better model. Future research should place attention on expanding the data basis in electricity price forecasting.

\noindent
\emph{Practical implications}: When developing pricing models, practitioners can achieve a reasonable performance with a simple model (\eg Seasonal-ARMA) that is built upon a wide range of predictors. 
\vspace{0.5cm}
\end{abstract}

%

\begin{keyword}

Electricity prices \sep forecasting \sep time series \sep machine learning \sep sensitivity analysis \sep machine learning

\emph{Article classification:} Research Paper 

\emph{Declaration of funding / declaration of interest:} none
\end{keyword}

\end{frontmatter}



\section{Introduction}


Participating in electricity markets requires decision support through accurate forecasts of future prices \citep{Nogales.2002,Weron.2014}. This is relevant for various activities; for example, producers can develop bidding strategies that maximize their financial return, while consumers aim at minimizing their spending \citep{Contreras.2003,Morales.2014}. {Analogously, it is important to hedge price risks, since electricity generation cannot always completely fulfill demand due to the physical characteristics of energy.}


These physical characteristics are one of the reasons for which electricity prices are known to be highly volatile \citep{Bierbrauer.2007,Weron.2007}. Another reason stems from seasonal patterns of electricity demand. In this regard, many electricity markets experience considerable changes in load depending on the time of year, day of the week and time of day \citep{Weron.2007}. At the same time, the supply side also contributes to large fluctuations in electricity prices, a result of the variable feed-ins from intermittent sources of power generation \citep{Green.2010,Wolff.2017,Woo.2011}. These feed-ins from renewable electricity sources are, in turn, highly dependent on weather conditions. Altogether, these stylized factors which jointly influence electricity prices, render price forecasting in energy markets challenging.


The forecasting of electricity prices relies upon a variety of different models \citep{Domanski.2017,Weron.2014,Gurtler.2017}. One common approach builds upon models from time series analysis, which are based on purely historic values \citep[cf.][]{Aggarwal.2009,Li.2005,Weron.2007,Weron.2014}. In this regard, \citet{Keles.2013} stress that, \emph{\textquote{as the feed-in from wind energy is more and more of crucial importance in electricity price modeling, also financial models have to integrate this new uncertain parameter in the modeling approach}}. However, these works rarely incorporate predictors that reflect both demand-side \emph{and} supply-side characteristics. Among the exceptions are \citep{Bello.2013,Cruz.2011,Erni.2012,Kristiansen.2012}, which improve predictive power based on a combination of load and wind infeeds. Otherwise, feed-ins from renewables, such as wind and solar power, are predominantly used separately: \citet{Bello.2013} consider solar infeeds, while wind is addressed by \citep{Bello.2013,Cruz.2011,Erni.2012,Kristiansen.2012}. Further works specifically draw upon the total grid load \citep[e.\,g.][]{Bello.2013,Kristiansen.2012,Panapakidis.2016,Weron.2008}, but, again, largely to the exclusion of supply-side factors.


Theoretical arguments help explain the reasons for which a wide range of variables affect electricity prices \citep{Wolff.2017} and may thus be utilized as viable predictors of price changes. On the supply side, renewable energy sources contribute additional electricity with almost no marginal costs, thus reducing the overall price \citep{Sensfuss.2008}. On the demand side, a higher grid load coincides with a higher demand and thus higher prices. Fuel prices are critical as they determine the costs for power generation from fossil sources \citep{Bello.2013}. Furthermore, we include the exchange rate as fuels are frequently traded in \US dollars and the exchange rate might thus be correlated with production costs.


As discussed above, previous research has rarely investigated the predictive power of supply- and demand-side factors in forecasting electricity prices. Yet little is known about how the individual variables can contribute to the improvement of forecasts. To this end, we study the relevance of various predictors. In order to account for the demand side, we include feed-ins from wind and solar power, as well as grid load. We specifically go beyond earlier research and incorporate predictors concerning fuel costs and the economic situation \citep{Wolff.2017}. These have been found to explain movements of electricity prices, although their prognostic capabilities are not well understood. 

This paper aims at comparing different predictors stemming from supply-side (solar and wind power generation), demand-side, fuel-related and economic influences. For this reason, we implement a broad range of non-linear models from machine learning and draw upon the information-fusion-based sensitivity analysis in order to disentangle the respective relevance of each predictor. We further show that these external predictors altogether decrease root mean squared errors by up to \SI{21.96}{\percent}. A Diebold-Mariano test statistically proves that the forecasting accuracy of the proposed models is superior.


{This work has direct implications for research and practice. As shown in this paper, the performance gain from including more predictors might be larger than from a better model. Developing a forecasting model without predictors might miss considerable predictive potential.  Hence, future research should place attention on expanding the data basis in electricity price forecasting. Furthermore, machine learning -- despite its flexibility in modeling non-linearities -- might be challenged in the dynamic setting of electricity price forecasting. In practice, it might thus be beneficial to choose upon a simple model that it is fairly prone against overfitting. For instance, practitioners can achieve a reasonable performance with a simple model (e.\,g. Seasonal-ARMA) that is built upon a wide range of predictors.}


The remainder of this paper is structured as follows. Section~\ref{sec:related_work} provides a literature overview of publications forecasting electricity prices, in which the majority of models ignore external impacts. To address this research gap, Section~\ref{sec:modeling} utilizes models from two different streams, namely time series analysis and machine learning. Subsequently, Section~\ref{sec:data} introduces our empirical setting. Based on this, Section~\ref{sec:forecasting} evaluates the sensitivity of the overall forecasting accuracy to different predictors. Section~\ref{sec:discussion} discusses managerial and policy implications, followed by concluding remarks.

\section{Background}
\label{sec:related_work}

Previous research on forecasting electricity prices can loosely be categorized by (i)~the predicted price, (ii)~the predictive model and (iii)~the predictors, as detailed in the following.

\subsection{Predicted price variables}


Related works differ with respect to which prices are subject to  forecasting. Forecasts often address spot prices where the hourly price of the intraday or day-ahead market needs to be predicted. These predictions are especially applicable to the day-to-day operations of market participants who are concerned with managing power generation and hedging risks \citep[cf.][]{Weron.2014}. Besides hourly resolution, \citet{Ziel.2018} forecast quarter-hourly spot prices. Other approaches are concerned average daily prices \citep{Bello.2013,Huurman.2012}. Even other works forecast prices of future contracts that are sold several days, months, or even years before delivery. These predictions thus facilitate long-term risk management, derivative pricing and strategic planning. As our work addresses hourly forecasts, we focus on relevant works from that area in the following.

\subsection{Predictive models}


Predictive models are subject to considerable variation . Previous literature has not only utilized common methods from time series analysis and machine learning. {Both time series models and machine learning have inherent benefits: the former yields parsimonious model specifications that are oftentimes more robust against overfitting, while the latter can learn non-linear relationships from data and thus offers more flexibility \citep{Hastie.2013}. The actual choice is eventually determined in computational experiments.} We refer to \citet{MartinezAlvarez.2015} and \citet{Weron.2008} for a recent overview and only summarize key directions in the following. 

On the one hand, time series models represent a common choice due to their ease of application and high explanatory power. In these models, different autoregressive terms help in predicting future prices and, by determining a suitable choice of lags and time dummies, these models can adapt well to seasonal characteristics. For instance, autoregressive models are widely used as a benchmark against which other models are compared \citep{Misiorek.2006,Uniejewski.2016,Weron.2008}. In practice, these models can be adapted to handle moving-average terms \citep[e.\,g.][]{Contreras.2003,Cruz.2011,Huurman.2012,Kristiansen.2012} or external regressors \citep[e.\,g.][]{Contreras.2003,Misiorek.2006,ECIS.2014}. Further extensions entail, for example, vector error correction models \citep{Bello.2013}, regime-switching variations \citep{Kosater.2006,Misiorek.2006,Weron.2008}, volatility clustering \citep{Erni.2012,Knittel.2005,Misiorek.2006} or transfer functions \citep{Nogales.2002,Nogales.2006}. Alternatively, \citet{Xu.2004} apply a wavelet transformation to hourly day-ahead prices by utilizing load as an external predictor. Yet autoregressive models from time series analysis are inherently limited to linear relationships \citep{Gurtler.2018}.


On the other hand, machine learning allows to address non-linear dependencies and interactions among predictors \citep{Hastie.2013}. Owing to this, machine learning has become popular for making data-driven forecasts in various areas of operations management. {Example use cases include sales forecasts in inventory management \citep[\eg,][]{SALES} and forecasts of stock prices \citep[\eg][]{FINANCE} or macroeconomic indicators \citep{MACRO}.} In the context of electricity price forecasting, different predictive models have been employed, for instance, neural networks \citep{Cruz.2011,Panapakidis.2016,Szkuta.1999}, regularization \citep{Ludwig.2015,Uniejewski.2016}, and tree-based approaches \citep{ECIS.2014,Ludwig.2015}. These largely achieve a lower forecast error. For instance, \citet{Ludwig.2015} suggest a regularized model and a tree ensemble as a means of implicit variable selection, where weather forecasts are taken into account. 

We note further observations as follows: re-estimating model parameters can sometimes help in predicting variables subject to seasonal patterns. For instance, multiple works \citep{Chen.2014,Contreras.2003,Voronin.2013,Weron.2008} recalibrate their model parameters for each trading day. However, the majority of aforementioned papers fail to utilize this performance improvement. Finally, \citet{Huurman.2012} point out that most studies only report in-sample errors, and do not evaluate the out-of-sample predictive performance.

\subsection{Predictors}


External predictors have recently been introduced to electricity price forecasting \citep[cf.][]{Keles.2013}. This approach is driven by the hypothesis that price movements are largely caused by supply and demand. However, we are aware of only a few studies that actually deal with such models, an observation that is confirmed by \citet{ECIS.2014} and \citet{Weron.2014}. Typical predictors quantify demand and supply factors. Further predictors are the prices of coal, oil, \etc \citep{Bello.2013}, system power reserves \citep{Szkuta.1999} and the price of emissions allowances \citep{Erni.2012}. Yet these works focus primarily on single predictors but do not compare their contribution to the overall forecast accuracy. 

It is the objective of this paper to advance the data-driven forecasting of electricity prices. For this purpose, we compare models that forecast hourly day-ahead electricity prices with external predictors. As our primary contribution and different from earlier work \citep{ECIS.2014}, we specifically evaluate the prognostic power of these predictors in forecasting electricity prices via an information-fusion-based sensitivity analysis. Moreover, we assess further economic variables and fuel costs. In addition, we conduct a rigorous evaluation of out-of-sample errors across a multiple year horizon. Based on this, we identify and measure the beneficial effects of incorporating externals based on the Diebold-Mariano test. Consistent with prior research, we allow for time-dependent parameters and utilize a moving window of training data in order to study the rolling re-estimation of coefficients.

\section{Models and procedures for forecasting electricity prices}
\label{sec:modeling} 

This section describes our routines for forecasting electricity prices. These can be grouped into (a)~a simple lag benchmark; (b)~time series models with external predictors; and (c)~approaches from machine learning. 

We introduce the following notation: let $p_t$ with $t = 1, \ldots, T$ denote the time series of electricity prices. We then aim at forecasting a future price denoted by $\tilde{p}_t$. We further utilize dummies $D_t^{(i)}$ that reflect seasonal effects (such as Saturday/Sunday). Finally, the variables $x_t^{(1)}$, \ldots, $x_t^{(k)}$ refer to $k$ external time series that work as additional predictors.

\subsection{Benchmarks}


We first test whether our models outperform a simple prediction from a historic lag. Hence we initialize our forecast by $\tilde{p}_t\define p_{t-168}$; \ie the historic price 7 days prior to time $t$. This simple approach has been proposed as a baseline by previous work \citep{Conejo.2005}, since the strong seasonality of electricity prices already explains a large proportion of the intraday and intra-week variations. This presents a frequent benchmark in electricity price forecasting \citep{Misiorek.2006}.

As a second benchmark, we utilize an autoregressive moving-average~(ARMA) model \cite[cf.][]{Luetkepohl.2007}. It describes a stationary stochastic process via an auto-regression that combines past values and a moving-average that incorporates past errors. This work follows earlier suggestions \citep{Weron.2007} concerning the seasonality of electricity prices and draws upon additional lags $p_{t-24}$, $p_{t-48}$ and $p_{t-168}$, which correspond to prices \SI{24}{\hour}, \SI{48}{\hour} and \SI{1}{week} before the current point in time, respectively. We thus incorporate these explicit lags together with $p$ lags stemming from the $p$ direct previous hours (or their forecasts) and refer to this model adaption as a Seasonal-ARMA$(p,q)$. 

\subsection{Time series model with external predictors}

The previous ARMA model can be extended by a linear combination of predictor variables, yielding an ARMAX model \citep{Luetkepohl.2007}. Thereby, the ARMAX model depends no longer only on past values, but also demand- and supply-side variables, as well as the other predictors. Analogously, we refer to the model as a Seasonal-ARMAX$(p,q)$ when it includes the additional lags $p_{t-24}$, $p_{t-48}$ and $p_{t-168}$. When choosing a model specification that involves merely autoregressive terms, this is called a dynamic linear regression or DLR for short \citep{Nogales.2002,Weron.2007}. Sometimes this model is also termed a transfer function or ARX process.

\subsection{Machine learning}

In addition to the aforementioned models, we utilize different approaches from machine learning. These learn a function $f$ parametrized by a vector $\omega$ from the training data to make predictions for new data. Consistent with previous models, the input consists of lagged prices $p_{t-24}$, $p_{t-48}$, $p_{t-168}$, dummy variables, and further predictors $x_t^{(i)}$, $i = 1, \ldots, k$. We draw upon the following non-linear models that have demonstrated high predictive power in a variety of tasks \citep{Bishop.2009,Hastie.2013}: support vector regression~(SVR), random forest~(RF), LASSO, ridge regression, principal component regression~(PCR), boosted linear model~(BLM). A detailed description is found in \ref{appendix:ML_models}. In addition to the aforementioned models, we train a linear ensemble based on the all of the considered models. That is, we fit a linear regression model to the predictions from each model obtained during cross-validation.

\subsection{Forecasting framework}
\label{sec:forecasting_framework}

\subsubsection{Rolling re-estimation}

This section presents the forecasting framework to predict electricity prices. Since our aim is to compare the above models in terms of their forecasting accuracy, we need to calculate the out-of-sample error. In other words, instead of measuring the model fit by using historic data, we need to estimate the model parameters with training data and, subsequently, evaluate our models using different testing data.  

When estimating the parameters of our models, we can choose between two options: (i)~coefficients can be determined once based on a training set, and then remain fixed throughout the whole testing set. Alternatively, (ii)~coefficients can be re-estimated daily for each forecast using a moving window of training data. Each re-estimation takes a new training set into account, shifted by one day. This idea is referred to as a \emph{rolling re-estimation} of parameters \citep[c.\,f.][]{Hu.1999,Li.2012}. That is, we re-estimate the models based on past data so as to better adapt to changes in the market structure, as well as the potentially variable influences of external predictors. As noted in Section~\ref{sec:related_work}, related publications seldom adopt the concept of re-estimating model parameters for every forecast. In this work, we utilize a rolling re-estimation at a daily level.

\subsubsection{Model tuning}
\label{sec:model_tuning}

In order to tune the hyper-parameters for the machine learning methods, we utilize time-series cross-validation that employs a rolling forecasting origin~\citep{Hyndman.2014}. Algorithm~\ref{alg:tuning} describes the tuning in order to find the best-performing parameters based on given training data indexed by $\mathcal{T}\subset\{1,\ldots,T\}$ and a subset size $d$. For this purpose, we define $\kappa=\frac{\abs{\mathcal T}-d}{24}$ subsets $\mathcal{T}_1, \ldots, \mathcal{T}_k$ of size $d$ that are chronologically ordered, each shifted by one day. Afterwards, we set the tuning range for each parameter and iterate over all possible combinations. In each iteration $i$, we train the model on each subset $\mathcal{T}_j$, $j = 1,\dots,k$, and evaluate the performance of the model forecast for the next 24 hours. For each parameter combination, the performance is measured as the average performance over all subsets. Finally, we return the  parameter setting corresponding to the best average performance. 

\begin{algorithm}[H]
\caption{Parameter tuning with time-series cross-validation}
\label{alg:tuning}
\footnotesize
\begin{algorithmic}
\State $\mathcal{T}\gets$ Training data which is chronologically ordered
\State $d\gets$ Subset size 
\State Define $k$ subsets of size $d$ $\mathcal{T}_1, \ldots, \mathcal{T}_\kappa$ maintaining the chronological order such that $\bigcup_{j=1}^\kappa\mathcal T_j =\mathcal T$.
\State $\mathcal{R}_{i}\gets$ Ranges of tuning parameter $i$ with $i = 1, \ldots, l$
\For{$(p_1, \ldots, p_l)$ \textbf{in} $\mathcal{R}_1 \times \ldots \times \mathcal{R}_l$}
  \For{$j$ \textbf{in} $1,\ldots,\kappa$}
    \State Train model $m$ with data from subset $\mathcal{T}_j$
    \State $\mathit{perf}_j\gets$ Performance of model $m$ for $p_t$ with $t\in\{\tilde{t}\mid\tilde{t}=\max\{\mathcal T_j\}+1,\ldots,24\}$.
  \EndFor
  \State $\mathit{perf}_{(p_1, \ldots, p_l)} \gets \frac{1}{\kappa}\sum\limits_{j = 1}^\kappa \mathit{perf}_j$
\EndFor 
\State \textbf{return} $\argmax\limits_{(p_1, \ldots, p_l)}{\;\mathit{perf}_{(p_1, \ldots, p_l)}}$
\end{algorithmic}
\end{algorithm}

In all of our models from machine learning, we set $d=504$ and $\abs{\mathcal T}=672$, which corresponds to \SI{3}{weeks} and \SI{4}{weeks}, respectively. We found that this choice yielded a suitable trade-off between providing sufficient data for training, but without incorporating observations from other seasons. The following hyper-parameters are used to train the models from machine learning:
\begin{itemize}
\item \emph{Support Vector Regression.} We use a $\varepsilon$-SVR with a Gaussian radial basis function as a kernel. The regularization constant $C$ is chosen from the set $\{2^{-1} , 2^1 , 2^3 , 2^5 , 2^6\}$ via cross validation, while the scale parameter $\sigma$ is estimated via the heuristic in \citep{Caputo.2002}.
\item \emph{Random Forest.} For each random forest model, we grow a total of $n = 500$ separate decision trees and fix the minimal node size to $5$. Finally, the optimal number of variables $m$ considered at each split is chosen from the set $\{\lfloor\sqrt{p}\rfloor-1,\lfloor\sqrt{p}\rfloor,\lfloor\sqrt{p}\rfloor+1\}$ via time-series cross-validation, where $p$ is the number of predictors.
\item \emph{LASSO and ridge estimator.} For both models, the optimal $\lambda$ is set via time-series cross-validation. The considered sets are constructed as follows. First, $\lambda_{max}$ is calculated, which is defined as the smallest $\lambda$ such that $\argmin_{\omega}J^q(\omega)=0$. Finally, the set is defined as an exponentially decreasing series ranging from $\lambda_{max}$ to $0.01\times\lambda_{max}$ of length $100$.
\item \emph{Principal component regression} The optimal number of principal components $k$ is found via time-series cross-validation from the set $\{1,\ldots,p\}$.
\item \emph{Boosted linear model.} For boosted linear models, one must set the appropriate number of boosting iterations $m_{\mathup{stops}}$. We compare the model performance for $m_{\mathup{stop}}\in\{500,1000,1500,2000,2500\}$ with $\nu$ held constant at $\frac{1}{2}$.
\end{itemize}
Regarding the Seasonal-ARMA(X) models, we proceed by choosing the model parameters (\ie the number $p$ of auto-regressive and the number $q$ of moving-average terms) as follows. On the one hand, we keep $p$ and $q$ fixed and, in each re-estimation, we always select a Seasonal-$\mathup{ARMAX}(1,1)$ model. On the other hand, we compare models with varying parameters $p$ and $q$. Here we choose the optimal combination of $p$ and $q$  in terms of the Akaike Information Criterion, corrected for finite samples in each re-estimation of a Seasonal-$\mathup{ARMAX}(p,q)$ process.

\subsubsection{Forecasting performance}
\label{sec:forecasting_accuracy}

We measure the predictive power of all models based on the following metrics, comparing the predicted price $\tilde{p}_t$ with true market prices $p_t$ after these have become available. As in earlier works \citep[see e.\,g.][]{Conejo.2005,Misiorek.2006}, we compare the mean absolute error~(MAE), the root mean squared error~(RMSE) and additional derived metrics that aggregate these on a daily and weekly basis.\footnote{The MAE and RMSE are defined via 
$\mathup{MAE} =  \frac{1}{T} \sum\limits_{t=1}^{T}{\abs{p_{t}-\tilde{p}_{t}}}$ and $\mathup{RMSE} = \sqrt{\frac{1}{T} \sum\limits_{t=1}^{T}{ {\left(p_{t} - \tilde{p}_{t}\right)}^{2}}}$.} Here literature \citep{Weron.2007} defines the daily root mean squared error~(DRMSE) and weekly root mean squared error~(WRMSE) as
\begin{equation}
\mathup{DRMSE} = \sqrt{\frac{1}{24} \sum\limits_{h=1}^{24}{(p_{h} - \tilde{p}_{h})^2}}
\sand
\mathup{WRMSE} = \sqrt{\frac{1}{168} \sum\limits_{h=1}^{168}{(p_{h} - \tilde{p}_{h})^2}} .
\end{equation}

\subsection{Information-fusion-based predictor sensitivity}

We implement an \emph{information-fusion-based} sensitivity analysis to compare the importance of the predictors used during model building \citep{Oztekin.2013}. For a model $m$ and a predictor $\alpha$, let $\mathup{RMSE}_m^{\alpha}$ denote the root mean squared error of model $m$ when fitted to the data using all predictors except $\alpha$. Conversely, $\mathup{RMSE}_m^0$ refers to the RMSE of the model fit with the full set of predictors. We then define the sensitivity of model $m$ to predictor $\alpha$ as 
\begin{equation}
\mathcal{S}_{m\alpha} = \frac{\mathup{RMSE}_m^{\alpha}}{\mathup{RMSE}_m^0}.
\end{equation}
This definition follows our intuition in the sense that we expect the RMSE for model $m$ to increase if an important predictor is removed and vice versa. We can then combine -- or \textquote{fuse} -- the information across a set of models $\mathcal M$ to derive the importance of predictor $\alpha$ as 
\begin{equation}
\mathcal S_{\alpha} = \sum_{m\in\mathcal{M}}\omega_m\mathcal{S}_{m\alpha},
\end{equation}
where $\omega_m$ is defined as the normalized $R^2$ value for model $m$. This allows us to later rank the importance of different variables for making accurate predictions.

\section{Data}
\label{sec:data}


In the subsequent evaluation, we use the following datasets. First of all, electricity prices originate from the so-called European Power Exchange (EPEX SPOT). The EPEX operates both intraday and day-ahead markets for Germany, Austria, France and Switzerland. We consider both hourly day-ahead spot and intraday prices on the German and Austrian market of the European Power Exchange from January 1, 2010 to December 31, 2014. This gives a total of \num{43824} observations each for day-ahead spot and intraday prices.

\begin{figure}[H]
	\centering
	\includegraphics[scale=0.5]{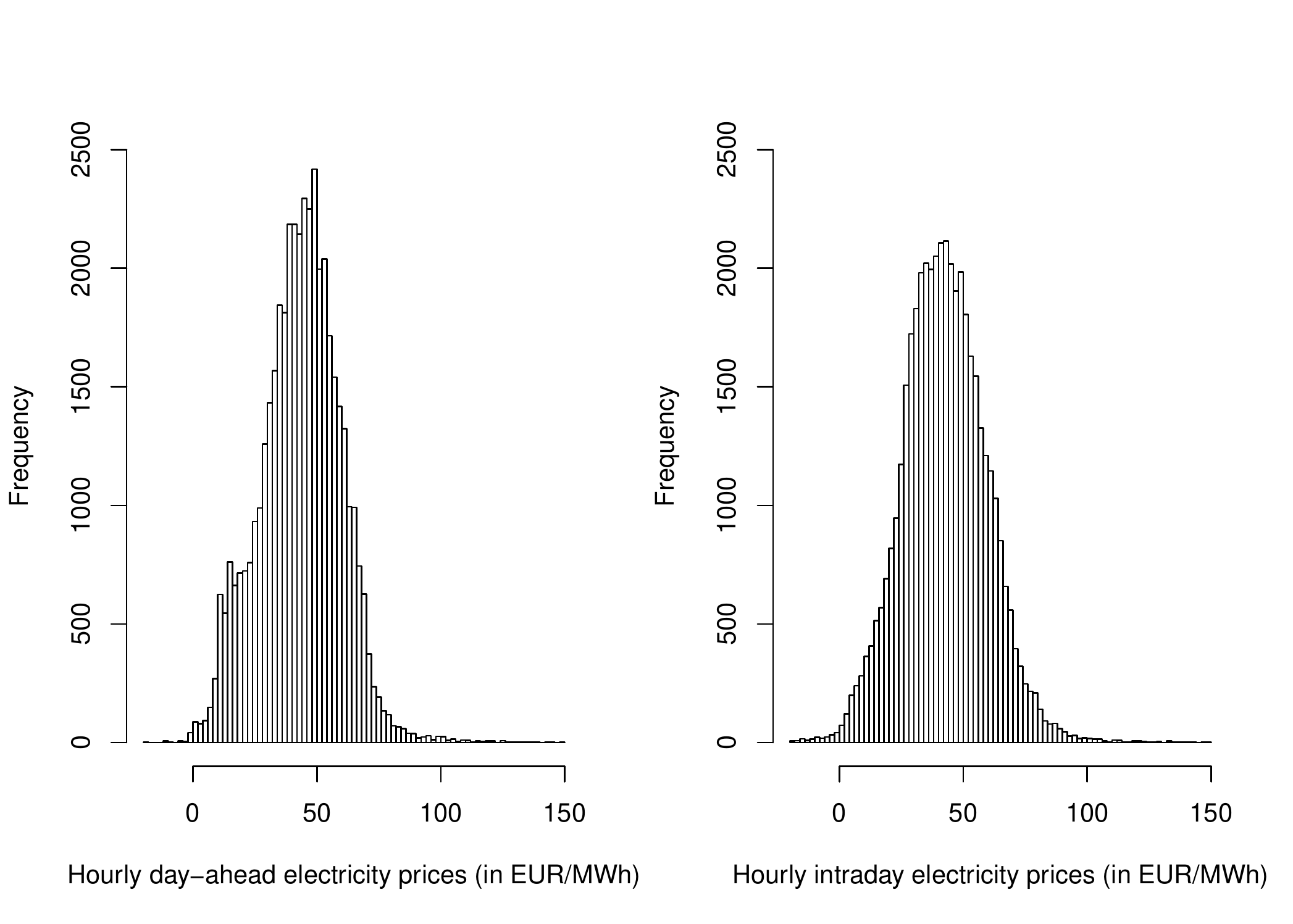}	
	\caption{Histogram of hourly day-ahead (left) and intraday (right) electricity prices, cropped to the range $[-20,150]$, for Germany at EPEX SPOT from January 1, 2010 to December 31, 2014 \citep{EPEX.2017}.}
	\label{fig:hourly_day-ahead_prices_histogram}
\end{figure}

The distribution of electricity prices is visualized in Figure~\ref{fig:hourly_day-ahead_prices_histogram}. This histogram indicates that prices are highly volatile. In particular, several positive and negative price spikes occur, mostly in December and January of each year. In addition, very high prices are encountered in February 2012 due to extreme weather conditions. According to Table~\ref{tbl:descriptive_statistics}, the mean price amounts to \SI{43.32}{\eur/MWh}, with a standard deviation of \SI{16.66}{\eur/MWh} for the day-ahead spot prices. For the intraday price, we observe a mean price of \SI{42.37}{\eur/MWh} and standard deviation of \SI{17.99}{\eur/MWh}. Day-ahead prices range from about \SI{-200.00}{\eur/MWh} to \SI{210.00}{\eur/MWh} and intraday prices from \SI{-270.11}{\eur/MWh} to \SI{272.95}{\eur/MWh}. Interestingly, the kurtosis of \num{5.52} and \num{9.61} for day-ahead and intraday prices are substantially higher than \num{3}, indicating the existence of heavy tails, which are probably caused by price spikes. Overall, we can deduce the following characteristics of the EPEX day-ahead and spot prices: high volatility, negative/positive price spikes, mean reversion and strong seasonality.

We utilize the following time series as external predictors (see Table~\ref{tbl:descriptive_statistics}): wind power generation, solar power generation and grid load. In many countries, regulatory publication requirements obligate transmission operators to release such volumes a day ahead. In Germany, this happens online via \citet{Transparency.2013}. We observe the following patterns: hourly wind infeeds range from about \SI{115}{MWh} to \SI{118212.4}{MWh}, with a mean of \SI{20622.87}{MWh}. Hourly solar infeeds range from \SI{0}{MWh} to \SI{96281.9}{MWh}, with a mean of \SI{10627.07}{MWh}. Both distributions are highly volatile and positively skewed, which indicates that wind generation and solar generation are often relatively low. In addition to renewables representing the supply side, we capture the demand side of electricity markets and thus incorporate the grid load. In addition, we determine fuel costs for electricity generation as represented by the prices for coal, gas and oil. We include the USD/euro exchange rate in our subsequent analysis to control for exchange rate effects pertaining to commodity trading.

\begin{center}
Insert Table~\ref{tbl:descriptive_statistics} about here
\end{center}

Table~\ref{tbl:correlation} lists the cross-correlation between electricity prices and predictor variables. While the electricity prices entail a correlation coefficient of \num{0.803}, the predictors show only a minor correlation among themselves, with the exception of the oil and gas price.\footnote{Renewables receive preferential treatment in Germany and thus their production causes prices and not vice versa.}

\begin{center}
Insert Table~\ref{tbl:correlation} about here
\end{center}

\section{Empirical results from forecasting}
\label{sec:forecasting}

We evaluate our proposed models in terms of their forecasting accuracy. All of these models differ in the set of included predictors. Thus, we introduce the following notation: let $p_{t-24}$, $p_{t-48}$ and $p_{t-168}$ denote lagged prices. Weekend dummies for Saturday and Sunday are given by $D_t^{\mathup{Sat}}$ and $D_t^{\mathup{Sun}}$ respectively, whereas $D_t^h$ represents $h = 1, \ldots, 24$ dummies for each hour. Finally, we include our predictors, representing supply-side, demand-side, fuel-related and economic variables. 

\subsection{Model comparison}


All predictive results for the day-ahead spot price are listed in Table~\ref{tbl:results_auction}. We now quantify the maximum gain in forecasting performance due to external variables. The best-performing baseline without external predictors is given by the SVR, which attains an RMSE of \num{9.25}. As a comparison, all models (excluding the ensemble) that utilize external predictors further improve the forecasting performance. The best-performing one overall is the Seasonal-$\textup{ARMAX}(p,q)$, which achieves an RMSE of \num{7.53}. This represents in an improvement of \SI{18.59}{\percent}. 


Table~\ref{tbl:results_intraday} lists the results for the intraday market. We summarize key findings in the following. Here the best-performing model without external variables is the Seasonal-$\textup{ARMA}(p,q)$, which attains an RMSE of \num{12.34}. Again, all predictive models with external predictors can outperform the best baseline without. Here the best-performing model with externals is again the Seasonal-$\textup{ARMAX}(p,q)$, with an RMSE of \num{9.63}, yielding an improvement of \SI{21.96}{\percent}.


Comparing the results for intraday and day-ahead spot prices, we note that the prediction accuracy for the latter is generally higher across all of the considered models. Evidently, external predictors contribute less to the formation of intraday prices than to day-ahead prices, as is also found in explanatory modeling \citep{Wolff.2017}, where both market design and lower trading volume are named as potential reasons.


We note that for both day-ahead and intraday spot prices, each model that includes additional predictors performs better than the corresponding model that excludes them. For example, the RMSE of the LASSO regression for intraday spot prices drops from \num{12.7} to \num{9.85} when changing from the benchmark model to the model with external predictors. Accordingly, we achieve an improvement of \SI{22.44}{\percent}. Similarly, the RMSE of the boosted linear model for day-ahead spot prices decreases from \num{9.50} to \num{7.79} -- a reduction of \SI{18.00}{\percent}. Overall, we see that all models substantially benefit from the inclusion of additional predictors.


Examining the ensemble models for both the intraday and day-ahead spot prices, we note that these perform worse than each individual model, both with and without external predictors. To shed light on the underlying reason, we analyzed the correlation between the values predicted by the different models for both day-ahead and intraday. These are all highly positively correlated (minimum correlation is \num{0.865} for predicted spot prices and \num{0.913} for day-ahead spot prices), which impairs the performance of the ensemble.


\begin{center}
Insert Table~\ref{tbl:results_auction} about here
\end{center}

\begin{center}
Insert Table~\ref{tbl:results_intraday} about here
\end{center}

\subsection{Statistical tests}

The Diebold-Mariano test \citep{Diebold.1995} provides statistical evidence that all predictive models with additional predictors (except those of the ensemble) yield more accurate forecasts. More precisely, we test the null hypothesis that the methods with these predictors are as accurate as models lacking these external inputs versus the alternative that the former offer greater prediction accuracy. In this context, prediction accuracy is measured with an $L_2$-norm as the loss function. Consistent with \citet{Bordignon.2013}, we specifically utilize a modified version of the Diebold-Mariano test for predicting a horizon of 24~hours at once. All test statistics are provided in Table~\ref{tbl:dm_auction} for the day-ahead spot prices and Table~\ref{tbl:dm_intraday} for the intraday spot prices. We note that in the case of both prices, all $P$-values of the Diebold-Mariano test for each model with external predictors are below the \SI{0.1}{\percent} significance threshold. The only exception is found in the ensemble model with external predictors, due to the poor performance of the model discussed earlier. Hence, in all cases, we can reject the null hypothesis at any common significance level and thus conclude that models including external predictors are more accurate than models without.

\begin{center}
Insert Table~\ref{tbl:dm_auction} about here
\end{center}

\begin{center}
Insert Table~\ref{tbl:dm_intraday} about here
\end{center}

\subsection{Predictor sensitivity analysis}

\begin{figure}[H]
\centering
\makebox[\textwidth]{
\begin{tabular}{cc}
\footnotesize (a) Predicted variable: day-ahead price & \footnotesize (b) Predicted variable: intraday price \\
\includegraphics[scale=0.45]{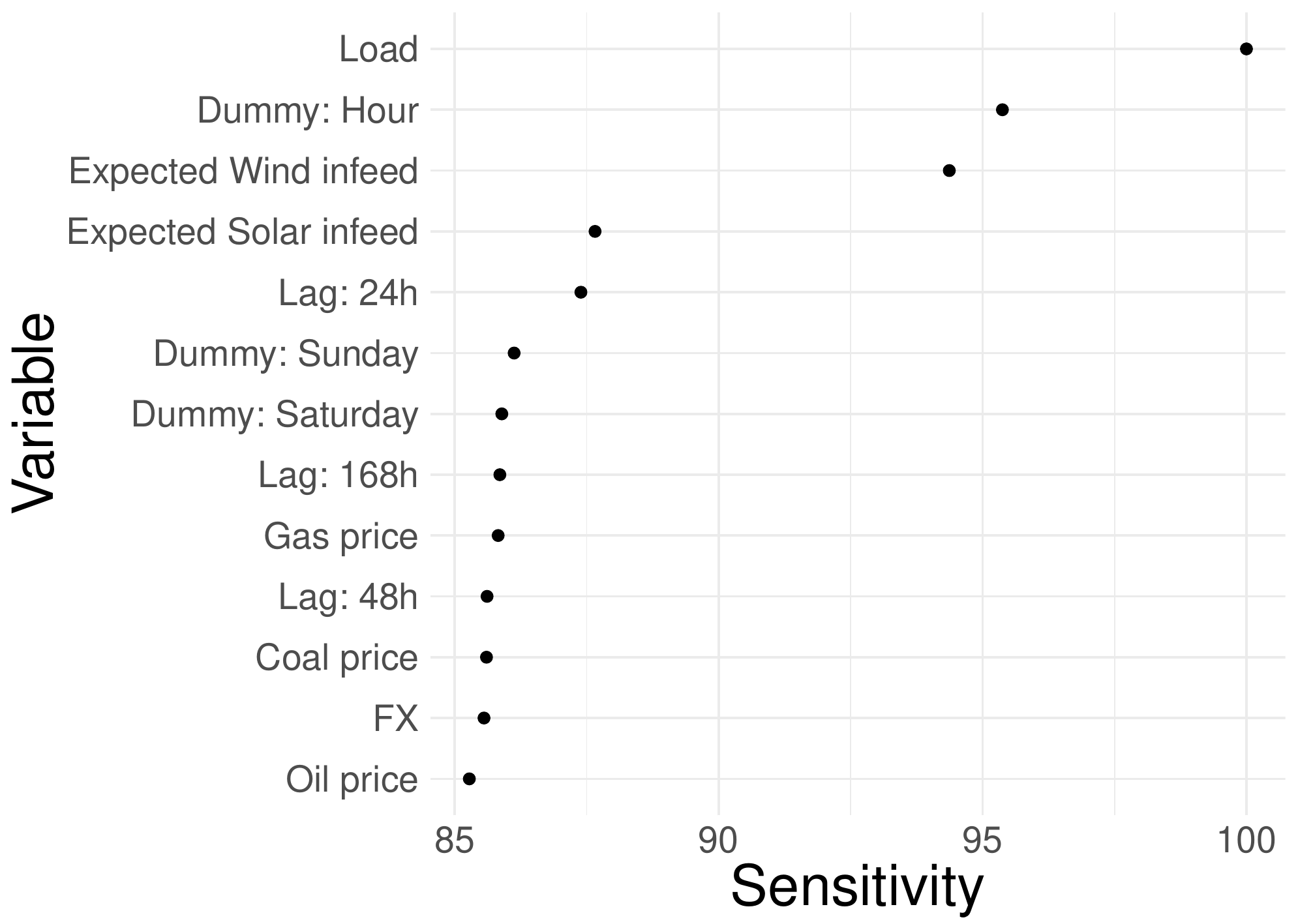} & \includegraphics[scale=0.45]{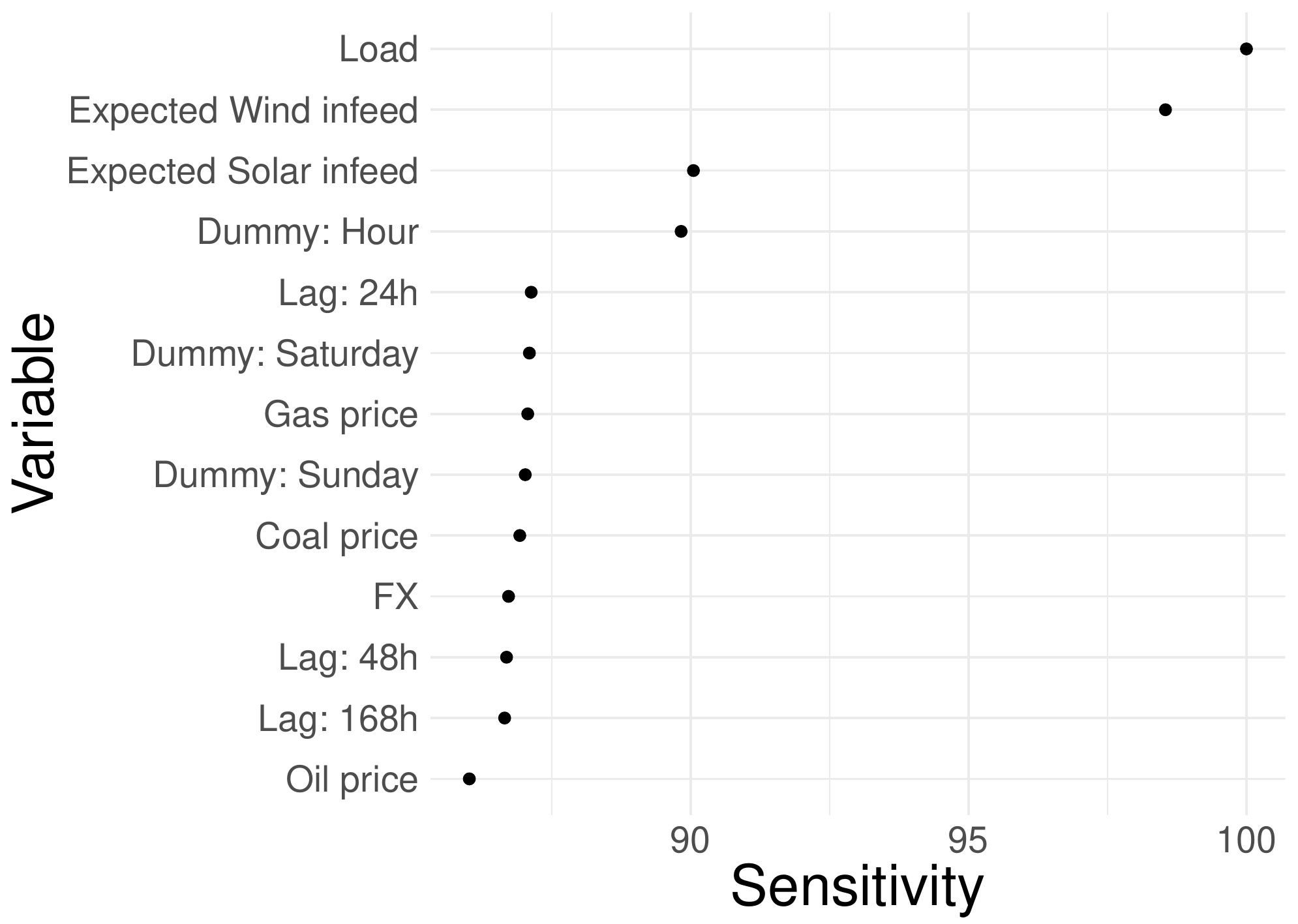}
\end{tabular}
}
\caption{Relevance (in \%) of the considered variables as measured by an information-fusion-based sensitivity analysis for the day-ahead~(left) and intraday~(right) spot price. For reasons of comparability across outcome variables, both metrics are scaled such that the highest relevance amounts to a relative sensitivity score of \SI{100}{\percent}.}
\label{fig:var_importance}
\end{figure}

We quantify the relative importance of the different predictors in computing the overall forecast. For this purpose, we draw upon the information-fusion-based sensitivity analysis. The corresponding results are reported in Figure~\ref{fig:var_importance}. Evidently, predictions for day-ahead~(left) and intraday~(right) spot prices reveal several differences. In both, load is ranked at the top. Supply-side factors (\ie wind and solar feed-ins) appear especially relevant for the intraday price. In contrast, day-ahead prices are more strongly driven by seasonal patterns (as given by the hourly dummy variables). We also observe a more relevant role for wind as compared to solar feed-ins. This is consistent with theoretical arguments \citep{Wolff.2017}. Finally, we observe fuel-related predictors to be ranked at the bottom. This matches our expectation, as gas turbines are predominantly added to the grid in times of electricity shortage. Coal-fired plants mostly contribute to the overall baseload and their operations are no particularly adaptable to short-term variations; hence, their production quota is covered by seasonal dummies rather than fuel costs. Oil-based power plants are rare in the EPEX market and thus affect prices with less intensity.

\section{Discussion}
\label{sec:discussion}

\subsection{Implications for practice and research}

Since the liberalization of electricity markets, prices for electricity are highly volatile. However, accurate price forecasts are invaluable when bidding in electricity markets to reduce the value-at-risk and minimize expenses. Hence, there is an interest in accurate price forecasts. To facilitate electricity price forecasting, this work compared the importance of different predictors. 

As shown in this paper, the performance gain from including additional predictors might be larger than that when developing a better forecasting model. Hence, practitioners can benefit by prioritizing models that are parsimonious yet that leverage the predictive utility from further variables. Put simply, focusing on forecasting models without any predictors (\ie models that are purely autoregressive) might miss considerable predictive potential. This should also be considered in future research where more attention on expanding the data basis around predictors is needed. By conducting this research, we guide practitioners in choosing effective predictors. Here we followed a theory-informed approach, that is, we categorized predictors into groups why they should theoretically act as determinants of electricity prices due to their influence on demand, supply, fuel costs, and the economy.

{This work utilized prediction models from two different areas, namely time series analysis and machine learning. Both have inherent strengths and weaknesses. Time series analysis with autoregressive structures yields parsimonious model formulations. This renders such models fairly robust against overfitting \citep{Hastie.2013}. Forecasting in the electricity market is especially prone to overfitting, since the production mix changes over the course of a year, thus also affecting coefficients that encode input-output relationships (\eg solar power might be a reliable source for electricity generation in summer, while it might has a tendency to serve as an excess supplier in winter). In contrast, machine learning models can learn various forms of non-linear relationships or interactions between predictors  \citep{Hastie.2013}. This gives them large flexibility to model arbitrary functions. Yet the larger parameter space (together with the need for hyperparameter tuning) also increases chances of overfitting, thus limiting the generalization power to out-of-sample data. Both streams find widespread application in practice (see Section~\ref{sec:related_work}), while the actual model choices must be determined in computational experiments. In line with this, we find that both linear and non-linear models appear suitable depending on the setting.}

The importance of predictors is likely to change over time. For instance, in many countries, the share of renewables remains subject to growth in the coming years and, hence, their feed-in volume will affect electricity markets accordingly \citep[e.\,g.][]{MacCormack.2010,Nieuwenhout.2013,Ryu.2010,SchleicherTappeser.2012}. As a consequence, one can expect an even stronger sensitivity of forecasts to supply-side predictors, thereby enhancing the accuracy of our data-driven models for electricity price forecasting. 

Future works might build upon this paper and experiment with additional innovations. For instance, the implementation of deep learning could yield additional improvements \citep{SALES}. From an explanatory point of view, a better understanding of the underlying mechanism (such as the potential exogeneity of predictors) might be a worthwhile subject for further research. {Risk management with regard to electricity trading requires not only point estimates of prices, but, to a growing extent, also interval forecasts \citep{Weron.2007}.} Hence, it would be beneficial to further evaluate to what extent the above models could be used for the purpose of risk management. 

\subsection{Policy implications}

As a policy implication, regulators should consider mechanisms that improves the accuracy of electricity price forecasting. The reasons behind this are manifold. For example, an enhanced understanding of price changes can allow electricity producers to control power plants more efficiently, thus reducing the pollution from greenhouse gases. Similarly, accurate forecasts are a necessary prerequisite for demand-side management, where they can help to mitigate price uncertainty \citep{ICIS.2012}.

The EPEX auction design requires participants to place bids for all delivery periods day-ahead due to the closing of the order book. Accordingly, they need to complete their predictions at the same time. The necessary values for seasonal dummies and historic dummies are available at that time. However, feed-ins, and even their expected volume, are published after the order book closes. Hence, this challenge could be addressed by appropriate regulatory publication requirements. Alternatively, market participants could build their own models for forecasting the predictor variables. For instance, they could predict expected feed-ins from renewables based on current weather conditions \citep[e.\,g.][]{Keles.2013}. Future research could specifically analyze how errors from noisy predictors propagate and evenutally affect electricity price forecasts.

\section{Conclusion}

Accurate forecasts of electricity prices are necessary for market participants producing and purchasing electricity. Many publications develop such forecasts, while the predictive power of external parameters has not yet been thoroughly compared. We thus study the predictive power of supply-side, demand-side, fuel-related and economic predictors. Based on this, our paper determines the sensitivity of electricity price forecasting to using different external predictors. When evaluating the out-of-sample errors, we find strong evidence that the inclusion of external predictors significantly improves forecasting accuracy. Depending on the chosen model, the root mean squared error drops by up to \SI{21.96}{\percent}, representing a significant increase in accuracy. Moreover, a Diebold-Mariano test provides statistical evidence that the predictors do indeed lead to superior performance. Overall, a sensitivity analysis of predictors ranks grid load as the most influential variable, followed by wind and solar power. 

\appendix

\section{Machine learning models}
\label{appendix:ML_models}

In the following, let 
\begin{equation}
\bm\gamma_t=\left(p_{t-24},p_{t-48},p_{t-168},x_t^{(1)},\ldots,x_t^{(k)},D_t^{(1)},\ldots,D_t^{(m)}\right)\in\mathbb{R}^p
\end{equation}
denote the combined vector of all predictors at time $t$. The $j$th components of a vector $c$ will be referred to by $c^{(j)}$ and $\mathcal T\subset\{1,\ldots T\}$ represents a set of consecutive indices. Then, the individual models are given as follows: 
\begin{itemize}
\item \emph{Support vector regression.} A classical support vector machine can be adapted for regression with a quantitative response yielding a so-called support vector regression~(SVR). The function $f$ is assumed to have the form $f(\bm\gamma_t)=h(\bm\gamma_t)^T\omega+\omega_0$, where $h(\gamma)$ is a fixed nonlinear mapping to an $m$-dimensional vector space. The mapping $h$ is not directly modeled but only the inner product $\langle h(\bm\gamma),h(\bm\gamma\prime)\rangle=\kappa(\bm\gamma,\bm\gamma\prime)$ with a kernel $\kappa$. Then, the parameter $\omega$ is estimated by minimizing 
\begin{equation}
J(w) = \sum\limits_{t\in\mathcal T}{L_\varepsilon\left(p_t-f(\bm\gamma_t)\right)} + C \norm{\omega^2}
\end{equation}
with a tuning parameter $C$. The $\varepsilon$-intensive error function
\begin{equation}
L_\varepsilon(r) = 
\begin{cases} 
0, & \text{if } \abs{r} < \varepsilon, \\
\abs{r} - \varepsilon, & \text{otherwise},
\end{cases}
\end{equation}
ignores errors of size less than $\varepsilon$. Analogously to support vector machines, the estimated model depends only on a subset of the training data because the error measure for building the model ignores any training data close (within a threshold $\varepsilon$) to the model prediction. By minimizing $J(w)$, one obtains a sparse set of coefficients $\alpha$ based on which we can make predictions for a new set of predictors $\tilde{\gamma}_t$ via
\begin{equation}
\tilde{p}_t = \hat{f}(\tilde{\bm\gamma}_t) = \omega_0 + \sum\limits_{i \in\mathcal T}\alpha_i \kappa(\tilde{\bm\gamma}_t, \bm\gamma_i). 
\end{equation}
\item \emph{Random forest.} Random forests \citep{Breiman.2001} are an ensemble learning method that constructs a large collection of $B$ de-correlated decision trees during training. For each $b=1,\ldots, B$, a bootstrap sample is drawn from the training set. Then, a decision tree $T_b(\bm\gamma,\omega)$ is grown to the bootstrap sample as follows: For each terminal node in the decision tree, $m\leq p$ of the predictors are randomly chosen. Among the $m$ predictors, the one which produces the split with the largest decrease in variance is chosen and the node is split into two new nodes. The procedure is repeated until a minimum node size is reached. The parameters $\omega$ are the split rules learned for each tree. After training, predicted values for unseen examples are calculated by averaging the predictors from all individual regression trees, \ie, $\tilde{p}_t=f(\bm\gamma_t) = \frac{1}{B}\sum_{b=1}^BT_b(\bm\gamma,\omega)$ 
\item \emph{LASSO and ridge estimator.} Both the least absolute shrinkage and selection operator~(LASSO) and ridge estimator assume a linear function $f$, \ie , $f(\bm\gamma_t)=\omega^T\bm\gamma_t+\omega_0$ for which the parameter $\omega$ is estimated via minimizing the function 
\begin{equation}
J^\alpha(\omega) = \sum_{t\in\mathcal T}\left(p_t-\omega_0-\sum_{j=1}^{p}\bm\gamma_{tj}\omega_j\right)^2+\lambda\sum_{j=1}^p\abs{\omega_j}^q
\end{equation} 
The LASSO estimator is obtained for $\alpha=1$ and the ridge estimator for $\alpha=2$. While both estimators are considered shrinkage estimators, only the LASSO performs model selection, as some coefficients will be exactly zero for a large enough $\lambda$.

\item \emph{Principal component regression.} Principal component regression~(PCR) is based on the idea of principal component analysis, which reduces dimensionality in datasets with many variables. Let $\tilde{\bm\Gamma}$ be the centered data matrix for predictors $\bm\gamma_t$ in the training set $\mathcal T$. Principal components are defined via the singular value decomposition of $\tilde{\bm\Gamma}=U\Sigma V^T$ with orthonormal matrices $U$ and $V$ and $\Sigma=\textup{diag}[\delta_1,\ldots,\delta_p]$ with $\delta_1\geq\ldots\geq\delta_p\geq0$. Then, for any $k\in\{1,\ldots,p\}$ the vector $z_k=\tilde{\bm\Gamma}v_k$ denotes the $k$th principal components. The principal component regression estimator for a given $k$ is then simply obtained by minimizing
\begin{equation}
	J(\omega)=\sum_{t\in\mathcal T}\left(p_t-\sum_{j=1}^{k}z_j\omega_j\right)^2.
\end{equation}

\item \emph{Boosted linear models.} For boosted linear models, a sequence of simple linear models is fit to the data as follows: For $m\in\{1,\ldots,m_{\mathup{stops}}\}$, we set 
\[\hat{f}^{[m]}(\bm\gamma_t) = \hat{f}^{[m-1]}(\bm\gamma_t) + \nu\hat{\omega}^{(j^*)}\bm\gamma_t^{(j^*)}, \]
where 
\[j^*=\argmin_{j\in\{1,\ldots,p\}}\sum_{t\in\mathcal T}\left[U_t-\hat{\omega}^{(j)}\bm\gamma_t^{(j)}\right]^2.\]
Here $U_t$ denotes the residual $p_t-\hat{f}^{[m-1]}(\bm\gamma_t)$, $\nu$ the so-called shrinkage parameter, and $\hat{f}^{(0)}=\frac{1}{\abs{\mathcal T}}\sum_{t\in\mathcal T}p_t$.
We note that the same predictor $\bm{\gamma}^{(j)}$ can be selected in multiple iterations. 
\end{itemize}



\bibliographystyle{model2-names-no-doi}\biboptions{authoryear}
\bibliography{literature}

\newpage 
\theendnotes
\newpage

\begin{table}
	\centering
\rotatebox{90}{
	
  \scriptsize
	\centering
	\input{descriptive_stats.tex}
}
	\caption{Descriptive statistics of electricity prices and the considered predictors.}
	\label{tbl:descriptive_statistics}
\end{table}

\begin{table}
\centering
\tiny
\begin{tabular}{l SSSSSSSS}
	\toprule
\input{corr_tbl.tex}
\end{tabular}
\caption{Cross-correlations between electricity prices and predictor variables.}
\label{tbl:correlation}
\end{table}

\begin{table}
	\centering
		\tiny
		\begin{tabular}{l ll SSSS }
			\toprule
			 \rottblhead{\tablehead Model}
			& \rottblhead{\tablehead Lagged Prices}
			& \rottblhead{\tablehead Dummies}
			%
			& \rottblhead{\tablehead RMSE}
			& \rottblhead{\tablehead MAE}
			& \rottblhead{\tablehead Avg. DRMSE}

			& \rottblhead{\tablehead Avg. WRMSE}

			\\
			\midrule
 \multicolumn{7}{c}{\textsc{Panel A: Benchmark models without external predictors}} \\ 
   \midrule			
			\input{auction_results.tex}
			\bottomrule
		\end{tabular}%
	\caption{Comparison of forecasting performance for the day-ahead spot price between January 01, 2010 and December 31, 2014. The best-performing model for each metric is highlighted in bold.}
	\label{tbl:results_auction}
\end{table}

\begin{table}
	\centering
		\tiny
		\begin{tabular}{l ll  SSSS }
			\toprule
			 \rottblhead{\tablehead Model}
			& \rottblhead{\tablehead Lagged Prices}
			& \rottblhead{\tablehead Dummies}
			%
			& \rottblhead{\tablehead RMSE}
			& \rottblhead{\tablehead MAE}
			& \rottblhead{\tablehead Avg. DRMSE}

			& \rottblhead{\tablehead Avg. WRMSE}

			\\
			\midrule
 \multicolumn{7}{c}{\textsc{Panel A: Benchmark models without external predictors}} \\ 
   \midrule
			\input{intraday_results.tex}
			\bottomrule
		\end{tabular}%
	\caption{Comparison of forecasting performance for the intraday spot price between January 01, 2010 and December 31, 2014. The best-performing model for each metric is highlighted in bold.}
	\label{tbl:results_intraday}
\end{table}

\begin{table}
	\centering
		\tiny
			\centering
			\makebox[\textwidth]{
		\begin{tabular}{l S  SSSSSSSSSS }
			\toprule
			&\multicolumn{11}{c}{Benchmark model without external predictors}\\
			\cmidrule(lr){2-12}
			\lcellt{Predictive model\\ with externals}
			& {\shortstack{Na{\"i}ve}}
			& {\shortstack{SVR}}
			& {\shortstack{DLR}}
			& {\shortstack{PCR}}
			%
			& {\shortstack{Ridge}}
			& {\shortstack{LASSO}}
			& {\shortstack{RF}}
			& {\shortstack{BLM}}
			& {ARMA(1,1)}
			& {ARMA($p$,$q$)}
			& {Ensemble}\\
			\midrule
			\input{auction_dm.tex}
			\bottomrule
			\multicolumn{12}{l}{\tiny{$^{***}p<0.001$, $^{**}p<0.01$, $^*p<0.05$}}
		\end{tabular}%
		}
	\caption{Test statistics of Diebold-Mariano test for the day-ahead spot price showing that the null hypothesis (methods with additional predictors are as accurate as models lacking these inputs in terms of RMSE) can be rejected for all cases. All corresponding $p$-values (except for the ensemble) are below the \SI{0.1}{\percent} significance threshold.}
	\label{tbl:dm_auction}
	
\end{table}
\begin{table}
	\centering
		\tiny
		\centering
		\makebox[\textwidth]{
		\begin{tabular}{l S  SSSSSSSSSS }
			\toprule
			&\multicolumn{11}{c}{Benchmark model without external predictors}\\
			\cmidrule(lr){2-12}
			\lcellt{Predictive model\\ with externals}
			& {\shortstack{Na{\"i}ve}}
			& {\shortstack{SVR}}
			& {\shortstack{DLR}}
			& {\shortstack{PCR}}
			%
			& {\shortstack{Ridge}}
			& {\shortstack{LASSO}}
			& {\shortstack{RF}}
			& {\shortstack{BLM}}
			& {ARMA(1,1)}
			& {ARMA($p$,$q$)}
			& {Ensemble}
			\\
			\midrule
			\input{intraday_dm.tex}
			\bottomrule
			\multicolumn{12}{l}{\tiny{$^{***}p<0.001$, $^{**}p<0.01$, $^*p<0.05$}}
		\end{tabular}%
		}
	\caption{Test statistics of the Diebold-Mariano test for the intraday spot price showing that the null hypothesis (methods with additional predictors are as accurate as models lacking these inputs in terms of RMSE) can be rejected for all cases. All corresponding $p$-values (except for the ensemble) are below the \SI{0.1}{\percent} significance threshold.}
	\label{tbl:dm_intraday}
\end{table}

\end{document}

%% file: descriptive_stats.tex
\begin{tabular}{lllSSSSSSr}
  \toprule
 & {\tablehead Unit } & {\tablehead Mean } & {\tablehead Median } & {\tablehead Min. } & {\tablehead Max. } & {\tablehead Std.\,dev. } & {\tablehead Skew. } & {\tablehead Kurt. } & {\tablehead Frequency } \\ 
  \midrule
Day-ahead spot price & \si{\eur/MWh} & 43.32 & 44.01 & -200.00 & 210.00 & 16.66 & -0.12 & 5.52 & Hourly \\ 
  Intraday spot price & \si{\eur/MWh} & 42.37 & 42.01 & -270.11 & 272.95 & 17.99 & -0.12 & 9.61 & Hourly \\ 
   \midrule
Wind infeed & \si{MWh} & 20622.87 & 14740.25 & 115.00 & 118212.40 & 18704.01 & 1.60 & 2.72 & Hourly \\ 
  Solar infeed & \si{MWh} & 10627.07 & 251.60 & 0.00 & 96281.90 & 18116.17 & 2.00 & 3.47 & Hourly \\ 
  Load & \si{MWh} & 55002.79 & 54910.50 & 29201.00 & 79884.00 & 10139.15 & -0.04 & -1.05 & Hourly \\ 
  Coal price & \si{\$/t} & 92.89 & 90.15 & 64.38 & 132.01 & 17.12 & 0.64 & -0.75 & Daily \\ 
  Gas price & \si{\eur/MWh} & 22.59 & 23.20 & 10.40 & 38.50 & 4.16 & -0.36 & 0.77 & Daily \\ 
  Oil price & \si{\$/bbl} & 102.08 & 108.01 & 58.69 & 126.62 & 14.85 & -0.85 & -0.31 & Daily \\ 
  Foreign exchange rate & \si{\$/\eur} & 1.33 & 1.33 & 1.19 & 1.49 & 0.06 & 0.08 & -0.40 & Daily \\ 
   \bottomrule
\end{tabular}

%% file: corr_tbl.tex
 & {\shortstack{Day-ahead\\spot price}} & {\shortstack{Intraday\\spot price}} & {\shortstack{Expected\\wind infeed}} & {\shortstack{Expected\\solar infeed}} & {Load} & {Coal price} & {Gas price} & {Oil price} \\ 
  \midrule
  {Intraday spot price} &  0.803 &  &  &  &  &  &  &  \\ 
  Expected wind infeed & -0.198 & -0.321 &  &  &  &  &  &   \\ 
  Expected solar infeed & -0.019 & -0.042 & -0.098 &  &  &  &  &   \\ 
  Load &  0.662 &  0.614 &  0.034 &  0.283 &  &  &  &   \\ 
  Coal price &  0.310 &  0.339 & -0.036 & -0.099 & -0.014 &  &  &    \\ 
  Gas price &  0.176 &  0.048 &  0.100 &  0.061 & -0.038 &  0.101 &  &    \\ 
  Oil price &  0.066 &  0.036 &  0.019 &  0.172 & -0.063 &  0.343 &  0.560 &    \\ 
  Foreign exchange rate &  0.108 &  0.112 &  0.011 & -0.025 &  0.075 &  0.406 & -0.162 &  0.340   \\ 
   \bottomrule

%% file: auction_results.tex
Na{\"i}ve & $p_{t-168}$ &  & 12.11 & 8.10 & 9.68  & 10.75  \\ 
  DLR & $p_{t-24}$, $p_{t-48}$, $p_{t-168}$ & $D_t^{Sat}$, $D_t^{Sun}$, $D_t^{h}$ & 9.55 & 6.62 & 7.96 &  8.72  \\ 
  Seasonal-ARMA(1,1) & $p_{t-24}$, $p_{t-48}$, $p_{t-168}$ & $D_t^{Sat}$, $D_t^{Sun}$, $D_t^{h}$ & 9.69 & 6.79 & 8.10  & 8.90  \\ 
  Seasonal-ARMA($p$,$q$) & $p_{t-24}$, $p_{t-48}$, $p_{t-168}$ & $D_t^{Sat}$, $D_t^{Sun}$, $D_t^{h}$ & 9.59 & 6.70 & 8.01  & 8.81 \\ 
  Ridge & $p_{t-24}$, $p_{t-48}$, $p_{t-168}$ & $D_t^{Sat}$, $D_t^{Sun}$, $D_t^{h}$ & 9.48 & 6.57 & 7.89  & 8.66  \\ 
  LASSO & $p_{t-24}$, $p_{t-48}$, $p_{t-168}$ & $D_t^{Sat}$, $D_t^{Sun}$, $D_t^{h}$ & 9.63 & 6.65 & 8.00  & 8.81  \\ 
  SVR & $p_{t-24}$, $p_{t-48}$, $p_{t-168}$ & $D_t^{Sat}$, $D_t^{Sun}$, $D_t^{h}$ & 9.25 & 6.27 & 7.61 &  8.37  \\ 
  PCR & $p_{t-24}$, $p_{t-48}$, $p_{t-168}$ & $D_t^{Sat}$, $D_t^{Sun}$, $D_t^{h}$ & 9.58 & 6.61 & 7.93 &  8.75  \\ 
  RF & $p_{t-24}$, $p_{t-48}$, $p_{t-168}$ & $D_t^{Sat}$, $D_t^{Sun}$, $D_t^{h}$ & 9.34 & 6.38 & 7.68 &  8.44  \\ 
  BLM & $p_{t-24}$, $p_{t-48}$, $p_{t-168}$ & $D_t^{Sat}$, $D_t^{Sun}$, $D_t^{h}$ & 9.50 & 6.55 & 7.88 & 8.67 \\ 
  Ensemble & $p_{t-24}$, $p_{t-48}$, $p_{t-168}$ & $D_t^{Sat}$, $D_t^{Sun}$, $D_t^{h}$ & 12.29 & 7.39 & 8.82  & 10.00  \\ 
   \midrule
\multicolumn{7}{c}{\textsc{Panel B: Predictive models utilizing external predictors}} \\
   \midrule
  DLR & $p_{t-24}$, $p_{t-48}$, $p_{t-168}$ & $D_t^{Sat}$, $D_t^{Sun}$, $D_t^{h}$ & 7.98 & 5.49 & 6.54   & 7.13  \\ 
  Seasonal-ARMAX(1,1) & $p_{t-24}$, $p_{t-48}$, $p_{t-168}$ & $D_t^{Sat}$, $D_t^{Sun}$, $D_t^{h}$ & 7.60 & 5.23 & 6.27 &  6.84  \\ 
  Seasonal-ARMAX($p$,$q$) & $p_{t-24}$, $p_{t-48}$, $p_{t-168}$ & $D_t^{Sat}$, $D_t^{Sun}$, $D_t^{h}$ & \bfseries 7.53 & \bfseries 5.19 & \bfseries 6.22 &  \bfseries 6.79  \\ 
  Ridge & $p_{t-24}$, $p_{t-48}$, $p_{t-168}$ & $D_t^{Sat}$, $D_t^{Sun}$, $D_t^{h}$ & 7.83 & 5.29 & 6.40  & 7.01  \\ 
  LASSO & $p_{t-24}$, $p_{t-48}$, $p_{t-168}$ & $D_t^{Sat}$, $D_t^{Sun}$, $D_t^{h}$ & 7.82 & 5.30 & 6.40  & 7.00  \\ 
SVR & $p_{t-24}$, $p_{t-48}$, $p_{t-168}$ & $D_t^{Sat}$, $D_t^{Sun}$, $D_t^{h}$ & 7.81 & \bfseries 5.19 & 6.36  & 6.97  \\ 
  PCR & $p_{t-24}$, $p_{t-48}$, $p_{t-168}$ & $D_t^{Sat}$, $D_t^{Sun}$, $D_t^{h}$ & 8.29 & 5.72 & 6.83 &  7.47  \\ 
  RF & $p_{t-24}$, $p_{t-48}$, $p_{t-168}$ & $D_t^{Sat}$, $D_t^{Sun}$, $D_t^{h}$ & 7.94 & 5.34 & 6.53 &  7.17  \\ 
  BLM & $p_{t-24}$, $p_{t-48}$, $p_{t-168}$ & $D_t^{Sat}$, $D_t^{Sun}$, $D_t^{h}$ & 7.79 & 5.33 & 6.39 &  6.97  \\ 
  Ensemble & $p_{t-24}$, $p_{t-48}$, $p_{t-168}$ & $D_t^{Sat}$, $D_t^{Sun}$, $D_t^{h}$ & 9.49 & 6.08 & 7.22 &  7.99  \\ 
  

%% file: intraday_results.tex
Na{\"i}ve & $p_{t-168}$ &  & 15.98 & 10.91 & 13.18 &  14.57 \\ 
  DLR & $p_{t-24}$, $p_{t-48}$, $p_{t-168}$ & $D_t^{Sat}$, $D_t^{Sun}$, $D_t^{h}$ & 12.63 & 8.70 & 10.49 &  11.49 \\ 
  Seasonal-ARMA(1,1) & $p_{t-24}$, $p_{t-48}$, $p_{t-168}$ & $D_t^{Sat}$, $D_t^{Sun}$, $D_t^{h}$ & 12.35 & 8.67 & 10.45  & 11.52  \\ 
  Seasonal-ARMA($p$,$q$) & $p_{t-24}$, $p_{t-48}$, $p_{t-168}$ & $D_t^{Sat}$, $D_t^{Sun}$, $D_t^{h}$ & 12.34 & 8.66 & 10.44  & 11.50  \\ 
  Ridge & $p_{t-24}$, $p_{t-48}$, $p_{t-168}$ & $D_t^{Sat}$, $D_t^{Sun}$, $D_t^{h}$ & 12.50 & 8.62 & 10.41 & 11.45  \\ 
  LASSO & $p_{t-24}$, $p_{t-48}$, $p_{t-168}$ & $D_t^{Sat}$, $D_t^{Sun}$, $D_t^{h}$ & 12.70 & 8.79 & 10.61 & 11.64 \\ 
  SVR & $p_{t-24}$, $p_{t-48}$, $p_{t-168}$ & $D_t^{Sat}$, $D_t^{Sun}$, $D_t^{h}$ & 12.41 & 8.43 & 10.25 &  11.30  \\ 
  PCR & $p_{t-24}$, $p_{t-48}$, $p_{t-168}$ & $D_t^{Sat}$, $D_t^{Sun}$, $D_t^{h}$ & 12.74 & 8.83 & 10.64 &  11.68  \\ 
  RF & $p_{t-24}$, $p_{t-48}$, $p_{t-168}$ & $D_t^{Sat}$, $D_t^{Sun}$, $D_t^{h}$ & 12.56 & 8.57 & 10.41 &  11.46  \\ 
  BLM & $p_{t-24}$, $p_{t-48}$, $p_{t-168}$ & $D_t^{Sat}$, $D_t^{Sun}$, $D_t^{h}$ & 12.56 & 8.62 & 10.42 & 11.43  \\ 
   \midrule
\multicolumn{7}{c}{\textsc{Panel B: Predictive models utilizing external predictors}} \\
   \midrule
  DLR & $p_{t-24}$, $p_{t-48}$, $p_{t-168}$ & $D_t^{Sat}$, $D_t^{Sun}$, $D_t^{h}$ & 10.38 & 7.18 & 8.60 &  9.38  \\ 
  Seasonal-ARMAX(1,1) & $p_{t-24}$, $p_{t-48}$, $p_{t-168}$ & $D_t^{Sat}$, $D_t^{Sun}$, $D_t^{h}$ & 9.65 & 6.70 & 8.11  & 8.84  \\ 
  Seasonal-ARMAX($p$,$q$) & $p_{t-24}$, $p_{t-48}$, $p_{t-168}$ & $D_t^{Sat}$, $D_t^{Sun}$, $D_t^{h}$ & \bfseries 9.63 & \bfseries 6.68 & \bfseries 8.09 &  \bfseries 8.81  \\ 
  Ridge & $p_{t-24}$, $p_{t-48}$, $p_{t-168}$ & $D_t^{Sat}$, $D_t^{Sun}$, $D_t^{h}$ & 10.18 & 6.89 & 8.38 &  9.20  \\ 
  LASSO & $p_{t-24}$, $p_{t-48}$, $p_{t-168}$ & $D_t^{Sat}$, $D_t^{Sun}$, $D_t^{h}$ & 9.85 & 6.68 & 8.14 &  8.92  \\ 
SVR & $p_{t-24}$, $p_{t-48}$, $p_{t-168}$ & $D_t^{Sat}$, $D_t^{Sun}$, $D_t^{h}$ & 10.49 & 6.99 & 8.58 &  9.50  \\ 
  PCR & $p_{t-24}$, $p_{t-48}$, $p_{t-168}$ & $D_t^{Sat}$, $D_t^{Sun}$, $D_t^{h}$ & 10.70 & 7.38 & 8.88  & 9.72  \\ 
  RF & $p_{t-24}$, $p_{t-48}$, $p_{t-168}$ & $D_t^{Sat}$, $D_t^{Sun}$, $D_t^{h}$ & 10.65 & 7.17 & 8.82 &  9.72  \\ 
  BLM & $p_{t-24}$, $p_{t-48}$, $p_{t-168}$ & $D_t^{Sat}$, $D_t^{Sun}$, $D_t^{h}$ & 9.99 & 6.85 & 8.26 &  9.01  \\ 
  Ensemble & $p_{t-24}$, $p_{t-48}$, $p_{t-168}$ & $D_t^{Sat}$, $D_t^{Sun}$, $D_t^{h}$ & 12.13 & 7.98 & 9.56 &  10.50  \\ 
  

%% file: auction_dm.tex
  DLR & 10.45\sym{***} & 6.99\sym{***} & 8.08\sym{***} & 7.54\sym{***} & 8.19\sym{***} & 8.66\sym{***} & 7.05\sym{***} & 7.62\sym{***} & 9.30\sym{***} & 8.72\sym{***} & 4.02\sym{***} \\ 
  Seasonal-ARMAX(1,1) & 11.16\sym{***} & 9.26\sym{***} & 9.78\sym{***} & 9.28\sym{***} & 10.16\sym{***} & 10.40\sym{***} & 8.92\sym{***} & 9.32\sym{***} & 11.28\sym{***} & 10.60\sym{***} & 4.25\sym{***} \\ 
  Seasonal-ARMAX($p$,$q$) & 11.36\sym{***} & 9.75\sym{***} & 10.33\sym{***} & 9.77\sym{***} & 10.75\sym{***} & 10.98\sym{***} & 9.39\sym{***} & 9.84\sym{***} & 11.80\sym{***} & 11.19\sym{***} & 4.31\sym{***} \\ 
  Ridge & 11.49\sym{***} & 10.27\sym{***} & 10.99\sym{***} & 10.26\sym{***} & 12.60\sym{***} & 12.30\sym{***} & 10.17\sym{***} & 10.43\sym{***} & 13.02\sym{***} & 11.94\sym{***} & 4.17\sym{***} \\ 
  LASSO & 11.39\sym{***} & 9.30\sym{***} & 10.28\sym{***} & 9.85\sym{***} & 11.12\sym{***} & 11.26\sym{***} & 9.19\sym{***} & 9.82\sym{***} & 11.82\sym{***} & 11.06\sym{***} & 4.16\sym{***} \\ 
 SVR & 11.25\sym{***} & 12.15\sym{***} & 10.49\sym{***} & 9.72\sym{***} & 12.06\sym{***} & 11.55\sym{***} & 10.44\sym{***} & 9.83\sym{***} & 14.09\sym{***} & 12.90\sym{***} & 4.14\sym{***} \\ 
  PCR & 10.43\sym{***} & 5.74\sym{***} & 7.18\sym{***} & 7.16\sym{***} & 7.56\sym{***} & 8.07\sym{***} & 6.05\sym{***} & 6.83\sym{***} & 8.28\sym{***} & 7.60\sym{***} & 3.83\sym{***} \\ 
  RF & 11.21\sym{***} & 10.41\sym{***} & 12.05\sym{***} & 11.14\sym{***} & 14.61\sym{***} & 14.29\sym{***} & 11.68\sym{***} & 11.50\sym{***} & 13.96\sym{***} & 13.34\sym{***} & 4.16\sym{***} \\ 
  BLM & 11.09\sym{***} & 8.42\sym{***} & 9.50\sym{***} & 8.97\sym{***} & 9.85\sym{***} & 10.26\sym{***} & 8.44\sym{***} & 9.05\sym{***} & 10.76\sym{***} & 10.16\sym{***} & 4.16\sym{***} \\ 
  Ensemble & 5.08\sym{***} & -0.51 & 0.15 & 0.19 & -0.02 & 0.34 & -0.32 & 0.03 & 0.45 & 0.25 & 3.13\sym{***} \\ 
  

%% file: intraday_dm.tex
  DLR & 12.04\sym{***} & 7.38\sym{***} & 9.01\sym{***} & 10.08\sym{***} & 9.06\sym{***} & 10.70\sym{***} & 9.20\sym{***} & 8.93\sym{***} & 7.95\sym{***} & 8.02\sym{***} & 6.70\sym{***} \\ 
  Seasonal-ARMAX(1,1) & 12.46\sym{***} & 8.37\sym{***} & 8.97\sym{***} & 10.90\sym{***} & 9.94\sym{***} & 10.49\sym{***} & 10.09\sym{***} & 8.93\sym{***} & 12.36\sym{***} & 12.36\sym{***} & 7.28\sym{***} \\ 
  Seasonal-ARMAX($p$,$q$) & 12.45\sym{***} & 8.36\sym{***} & 9.03\sym{***} & 10.90\sym{***} & 9.95\sym{***} & 10.53\sym{***} & 10.13\sym{***} & 8.99\sym{***} & 12.32\sym{***} & 12.33\sym{***} & 7.31\sym{***} \\ 
  Ridge & 12.59\sym{***} & 8.95\sym{***} & 10.69\sym{***} & 12.50\sym{***} & 11.53\sym{***} & 13.28\sym{***} & 11.53\sym{***} & 10.66\sym{***} & 9.65\sym{***} & 9.73\sym{***} & 7.03\sym{***} \\ 
  LASSO & 12.89\sym{***} & 10.11\sym{***} & 10.63\sym{***} & 13.35\sym{***} & 12.73\sym{***} & 13.49\sym{***} & 12.41\sym{***} & 10.58\sym{***} & 11.09\sym{***} & 11.18\sym{***} & 7.31\sym{***} \\ 
 SVR & 12.49\sym{***} & 10.54\sym{***} & 8.68\sym{***} & 11.91\sym{***} & 11.81\sym{***} & 11.70\sym{***} & 10.46\sym{***} & 8.49\sym{***} & 8.48\sym{***} & 8.61\sym{***} & 6.59\sym{***} \\ 
  PCR & 11.70\sym{***} & 6.64\sym{***} & 8.33\sym{***} & 10.01\sym{***} & 8.52\sym{***} & 10.26\sym{***} & 8.84\sym{***} & 8.27\sym{***} & 7.05\sym{***} & 7.10\sym{***} & 6.31\sym{***} \\ 
  RF & 11.99\sym{***} & 8.14\sym{***} & 8.29\sym{***} & 12.64\sym{***} & 11.56\sym{***} & 11.32\sym{***} & 12.16\sym{***} & 8.20\sym{***} & 8.81\sym{***} & 8.85\sym{***} & 6.37\sym{***} \\ 
  BLM & 12.75\sym{***} & 8.97\sym{***} & 10.48\sym{***} & 12.08\sym{***} & 11.11\sym{***} & 12.74\sym{***} & 11.16\sym{***} & 10.47\sym{***} & 9.83\sym{***} & 9.92\sym{***} & 7.21\sym{***} \\ 
  Ensemble & 7.58\sym{***} & 0.58 & 0.99 & 1.41 & 0.75 & 1.17 & 0.87 & 0.85 & 0.42 & 0.39 & 3.63\sym{***} \\ 
  